\documentclass[twocolumn,fleqn]{article}

\usepackage{amsmath}
\usepackage{amssymb} 
\usepackage{lastpage}
\usepackage{eso-pic}
\usepackage{hyperref}
\usepackage{booktabs}
\usepackage{gensymb}
\usepackage{tabularx}%
\usepackage{xcolor}%
\usepackage{siunitx}%
\usepackage{cleveref}
\usepackage{float}
\usepackage{acronym}
\usepackage{calc}
\usepackage[style=base]{caption}
\usepackage{subcaption}
\usepackage{microtype}
\usepackage{cite}
\usepackage{bm}
\usepackage{flushend}
\usepackage{tikz}
\usepackage{subcaption}
\usepackage{multirow}   

\usepackage{bm}

\usepackage[utf8]{inputenc}	
\usepackage[T1]{fontenc}   	
\usepackage{microtype}     	

\usepackage{authblk} 
\usepackage{morewrites} 
\usepackage[charter]{mathdesign}
\usepackage[a4paper, margin=2cm]{geometry} 

\acrodef{pca}[PCA]{Principle Component Analysis}
\acrodef{pc}[PC]{Principle Component}

\acrodef{dtft}[DTFT]{discrete-time Fourier transform}

\acrodef{oe}[OE]{occlusion effect}
\acrodef{bwe}[BWE]{bandwidth extension}

\acrodef{bc}[BC]{bone-conducted}
\acrodef{ac}[AC]{air-conducted}

\acrodef{nam}[NAM]{Non-Audible Murmur}

\acrodef{bcm}[BCM]{bone-conduction microphone}
\acrodef{acm}[ACM]{air-conduction microphone}

\acrodef{anc}[ANC]{active noise cancellation}
\acrodef{aoc}[AOC]{active occlusion cancellation}
\acrodef{ff-anc}[FF-ANC]{feedforward \ac{anc}}
\acrodef{ff}[FF]{feedforward}
\acrodef{fb}[FB]{feedback}

\acrodef{adc}[ADC]{analog-digital-converter}
\acrodef{dac}[DAC]{digital-analog-converter}

\acrodef{ir}[IR]{impulse response}

\acrodef{ps}[PS]{power spectrum}
\acrodefplural{ps}[PS]{power spectra}
\acrodef{psg}[PSG]{power spectral gain}
\acrodef{stpsg}[ST-PSG]{short-time \acl{psg}}
\acrodef{rms}[RMS]{root-mean-suared}
\acrodef{msc}[MSC]{magnitude-squared coherence}
\acrodef{snr}[SNR]{signal-to-noise ratio}

\acrodef{dsp}[DSP]{digital signal processing}

\acrodef{imc}[IMC]{internal model control}
\acrodef{sos}[SOS]{second-order sections}
\acrodef{tf}[TF]{transfer function}

\acrodef{psd}[PSD]{power spectral density}
\acrodefplural{psd}[PSDs]{power spectral densities}

\acrodef{asd}[ASD]{amplitude spectral density}
\acrodefplural{asd}[ASDs]{amplitude spectral densities}

\acrodef{gcd}[GCD]{great circle distance}

\acrodef{vad}[VAD]{voice activity detection}

\acrodef{ov}[OV]{own voice}

\acrodef{pg}[PSG]{power spectral gain}

\acrodef{sota}[SOTA]{state-of-the-art}

\acrodef{imu}[IMU]{inertial measurement unit}
\definecolor{rwthblue}{rgb}{0,   0.3294,    0.6235}
\definecolor{rwthred}{rgb}{0.8000,    0.0275,    0.1176}
\definecolor{rwthturquoise}{rgb}{0,    0.5961,    0.6314}
\definecolor{rwthblack}{rgb}{0,0,0}
\definecolor{rwthgreen}{rgb}{0.34118,0.67059,0.15294}%
\definecolor{rwthpurple}{rgb}{0.4784,0.4353,0.6745}
\definecolor{rwthyellow}{rgb}{0.96471,0.65882,0.00000}%
\definecolor{rwthmagenta}{rgb}{0.89020,0.00000,0.40000}%

\definecolor{rwthbluelight}{rgb}{0.412,0.629,0.824}
\definecolor{rwthredlight}{rgb}{1.000,0.517,0.574}
\definecolor{rwthturquoiselight}{rgb}{0.416,0.808,0.831}
\definecolor{rwthblacklight}{rgb}{0.200,0.200,0.200}
\definecolor{rwthgreenlight}{rgb}{0.657,0.871,0.535}
\definecolor{rwthorangelight}{rgb}{1.000,0.841,0.500}
\definecolor{rwthmagentalight}{rgb}{1.000,0.500,0.725}
\definecolor{rwthbordeauxlight}{rgb}{0.831,0.457,0.553}
\definecolor{rwthvioletlight}{rgb}{0.580,0.389,0.553}
\definecolor{rwthyellowlight}{rgb}{1.000,0.965,0.500}
\definecolor{rwthlightgreenlight}{rgb}{0.961,1.000,0.500}
\definecolor{rwthpurplelight}{rgb}{0.747,0.719,0.875}
\definecolor{rwthdarkgreenlight}{rgb}{0.298,0.584,0.596}

\definecolor{mycolor1}{rgb}{0.00000,0.28000,0.53000}%
\definecolor{mycolor2}{rgb}{0.00000,0.41176,0.77941}%
\definecolor{mycolor3}{rgb}{0.68000,0.02333,0.10000}%
\definecolor{mycolor4}{rgb}{1.00000,0.03431,0.14706}%

\definecolor{mycolor2_delay}{rgb}{0.03430,0.24386,1.00000}%
\definecolor{mycolor6_delay}{rgb}{0.92000,0.19414,0.03156}%
\definecolor{mycolor4_delay}{rgb}{0.82621,0.02881,0.84000}%

\definecolor{yellowarrow}{rgb}{0.95,0.75,0.1}%

\colorlet{singleblack}{rwthblacklight}
\colorlet{singlered}{rwthredlight}
\colorlet{singlemagenta}{rwthmagentalight}
\colorlet{singlegreen}{rwthgreenlight}
\colorlet{singleturquoise}{rwthturquoiselight}
\colorlet{singleblue}{rwthbluelight}
\colorlet{singlepurple}{rwthpurplelight}
\colorlet{singleyellow}{rwthyellowlight}

\colorlet{averageblack}{rwthblack}
\colorlet{averagered}{rwthred}
\colorlet{averagemagenta}{rwthmagenta}
\colorlet{averagegreen}{rwthgreen}
\colorlet{averageturquoise}{rwthturquoise}
\colorlet{averageblue}{rwthblue}
\colorlet{averagepurple}{rwthpurple}
\colorlet{averageyellow}{rwthyellow}

\definecolor{mygray}{HTML}{808080}

\usepackage{amsmath}

\renewcommand{\vec}[1]{\mathbf{#1}}
\newcommand{\norm}[1]{|\hspace{-0.03cm}| #1 |\hspace{-0.03cm}|}
\newcommand{\transposed}{^\text{T}}

\newcommand{\accelAxisIdx}{j}

\newcommand{\ith}{$n$th\ }

\newcommand{\headsetSigLabel}{s}
\newcommand{\headsetSig}{\headsetSigLabel(k)}

\newcommand{\accelSigIndividualLabel}{a}
\newcommand{\accelSigIthLabel}{\accelSigIndividualLabel_{\accelAxisIdx}}
\newcommand{\accelSigXLabel}{\accelSigIndividualLabel_{x}}
\newcommand{\accelSigYLabel}{\accelSigIndividualLabel_{y}}
\newcommand{\accelSigZLabel}{\accelSigIndividualLabel_{z}}
\newcommand{\accelSigIth}{\accelSigIthLabel(k)}
\newcommand{\accelSigX}{\accelSigXLabel(k)}
\newcommand{\accelSigY}{\accelSigYLabel(k)}
\newcommand{\accelSigZ}{\accelSigZLabel(k)}

\newcommand{\accelSigLabel}{\vec{\accelSigIndividualLabel}}
\newcommand{\accelSig}{\accelSigLabel(k)}
\newcommand{\accelSigVec}{\begin{bmatrix}
    \accelSigX & \accelSigY & \accelSigZ
\end{bmatrix}^\text{T}}

\newcommand{\zTrafVar}{f}

\newcommand{\tfHeadsetToAccelIthLabel}{H_{\headsetSigLabel\accelSigIthLabel}}\newcommand{\tfHeadsetToAccelXLabel}{H_{\headsetSigLabel\accelSigXLabel}}
\newcommand{\tfHeadsetToAccelYLabel}{H_{\headsetSigLabel\accelSigYLabel}}
\newcommand{\tfHeadsetToAccelZLabel}{H_{\headsetSigLabel\accelSigZLabel}}
\newcommand{\tfHeadsetToAccelIth}{\tfHeadsetToAccelIthLabel(\zTrafVar)}
\newcommand{\tfHeadsetToAccelX}{\tfHeadsetToAccelXLabel(\zTrafVar)}
\newcommand{\tfHeadsetToAccelY}{\tfHeadsetToAccelYLabel(\zTrafVar)}
\newcommand{\tfHeadsetToAccelZ}{\tfHeadsetToAccelZLabel(\zTrafVar)}

\newcommand{\tfHeadsetToAccelLabel}{\vec{H}_{s\vec{a}}}
\newcommand{\tfHeadsetToAccel}{\tfHeadsetToAccelLabel(\zTrafVar)}
\newcommand{\tfHeadsetToAccelVec}{\begin{bmatrix}
        \tfHeadsetToAccelX & \tfHeadsetToAccelY & \tfHeadsetToAccelZ
\end{bmatrix}^\text{T}}
\newcommand{\tfHeadsetToAccelNorm}{\norm{\vec{H}_{s\vec{a}}(f)}}

\newcommand{\accelSigIthSignalLabel}{v_{\accelAxisIdx}}
\newcommand{\accelSigIthSignal}{\accelSigIthSignalLabel(k)}
\newcommand{\accelSigIthNoiseLabel}{n_{\accelAxisIdx}}
\newcommand{\accelSigIthNoise}{\accelSigIthNoiseLabel(k)}

\newcommand{\psAccelIthSignalPlusNoise}{\Phi_{\smash{\accelSigIthLabel}}(f)}
\newcommand{\psAccelIthSignal}{\Phi_{\smash{\accelSigIthSignalLabel}}(f)}
\newcommand{\psAccelIthNoise}{\Phi_{\smash{\accelSigIthNoiseLabel}}(f)}
\newcommand{\psAccelIth}{\Phi_{\smash{\accelSigIthLabel}}(f)}

\newcommand{\asAccelIthSignal}{\sqrt{\psAccelIthSignal}}
\newcommand{\asAccelIthNoise}{\sqrt{\psAccelIthNoise}}

\newcommand{\psAccelSum}{\Phi_{\accelSigLabel}(f)}
\newcommand{\psAccelSumExpanded}{\sum_{\accelAxisIdx\in\{x,y,z\}}\psAccelIth}

\newcommand{\cpsHeadsetAccelIthSignal}{\Phi_{\accelSigIthLabel\headsetSigLabel}(f)}
\newcommand{\psHeadsetSignal}{\Phi_{\headsetSigLabel}(f)}

\newcommand{\pcIth}{\vec{a}_{\text{p}n}}
\newcommand{\pcFirst}{\vec{a}_{\text{p}1}}

\newcommand{\pcSecond}{\vec{a}_{\text{p}2}}
\newcommand{\pcThird}{\vec{a}_{\text{p}3}}

\newcommand{\varSym}{\sigma^2}

\newcommand{\varIth}{\varSym_{\pcIth}}

\newcommand{\pcFirstWorld}{\vec{w}_{\text{p}1}}

\newcommand{\pcMeanFirst}{\overline{\vec{a}}_{\text{p}1}}
\newcommand{\pcMeanSecond}{\overline{\vec{a}}_{\text{p}2}}
\newcommand{\pcMeanThird}{\overline{\vec{a}}_{\text{p}3}}

\newcommand{\sysAccel}{\mathcal{A}}
\newcommand{\sysHead}{\mathcal{H}}
\newcommand{\sysWorld}{\mathcal{W}}

\newcommand{\gravWorld}{\vec{w}_\text{g}}
\newcommand{\gravAccelOne}{\vec{a}_{\text{g}}^{\text{calib}}}
\newcommand{\gravAccelTwo}{\vec{a}_{\text{g}}^{\text{tilt}}}

\newcommand{\deltaRot}{\Delta \vec{R}}

\newcommand{\earbudOrientationWorld}{\vec{W}_{\sysAccel}}
\newcommand{\earbudOrientationWorldOne}{\vec{W}_{\sysAccel}^{\text{calib}}}

\newcommand{\headtrackerOrientation}{\vec{W}_{\sysHead}}
\newcommand{\headtrackerOrientationOne}{\vec{W}_{\sysHead}^{\text{calib}}}
\newcommand{\headtrackerOrientationTwo}{\vec{W}_{\sysHead}^{\text{tilt}}}

\newcommand{\direction}{\vec{d}}
\newcommand{\projSigLabel}{p_{\direction}}
\newcommand{\projSig}{\projSigLabel(k)}

\newcommand{\snr}{\text{SNR}(f)}

\newcommand{\psProjected}{\Phi_{\projSigLabel}(f)}

\newcommand{\psProjPcFirst}{\Phi_{p_{\pcMeanFirst}}\hspace{-0.3em}(f)}
\newcommand{\psProjPcThird}{\Phi_{p_{\pcMeanThird}}\hspace{-0.3em}(f)}

\newcommand{\pGain}{G(f,\direction)}

\newcommand{\pGainAveragedLabel}{\overline{G}}
\newcommand{\pGainAveraged}{\pGainAveragedLabel(\direction)}

\newcommand{\directionMax}{\direction_\text{max}}

\newcommand{\directionOpt}{\vec{d}_\text{opt}}

\newcommand{\earbudOrientationAveraged}{\overline{\vec{W}}_{\sysAccel}}

\newcommand{\pGainModified}{G(f,\direction_1,\direction_2)}

\newcommand{\dSet}{\mathcal{D}}

\newcommand{\dSetAlpha}{\mathcal{C}(\alpha)}

\newcommand{\pGainAveragedNormalizedLabel}{\overline{G}_\text{n}}
\newcommand{\pGainAveragedNormalized}{\pGainAveragedNormalizedLabel(\direction)}

\newcommand{\pGainAveragedMaxAlpha}{\pGainAveragedNormalizedLabel^\text{max}(\alpha)}
\newcommand{\pGainAveragedMinAlpha}{\pGainAveragedNormalizedLabel^\text{min}(\alpha)}

\DeclareSIUnit\grav{g}
\date{}

\usepackage{pifont}

\usepackage{environ}

\usepackage{titlesec}
\titleformat{\section}{\large\bfseries}{\thesection}{0.33em}{}
\titleformat{\subsection}{\normalsize\bfseries}{\thesubsection}{0.33em}{}
\titleformat{\subsubsection}{\normalsize\bfseries}{\thesubsubsection}{0.33em}{}

\usepackage{fancyhdr}

\newcommand{\preprintfooter}{%
    \footnotesize Preprint -- revised version published in \href{https://doi.org/10.17743/jaes.2026.0301}{\emph{J.~Audio Eng.\ Soc.}, vol.~74, no.~11, pp.~796--810, 2026}%
}

\fancypagestyle{plain}{%
  \fancyhf{}%
  \fancyfoot[C]{\preprintfooter}%
  \fancyfoot[R]{\thepage}%
}

\begin{document}

\title{\Huge Sensing Bone-Conducted Speech with Earbuds}

\author{Christoph Weyer}
\author{Peter Jax}
\affil{Institute of Communication Systems (IKS), RWTH Aachen University, Aachen, Germany}
\affil{\texttt{\{\href{mailto:weyer@iks.rwth-aachen.de}{weyer}, \href{mailto:jax@iks.rwth-aachen.de}{jax}\}@iks.rwth-aachen.de}}

\twocolumn[
  \begin{@twocolumnfalse}
  \maketitle
    \begin{abstract}
    Clear capture of the wearer's own voice (OV) is essential when using earbuds for mobile communication. However, OV capture remains challenging in noisy environments. Bone-conducted (BC) speech, which can be sensed as vibrations of the earbud housing, can be used to improve OV capture. However, neither bandwidth nor spatial characteristics of OV-induced earbud vibrations have been analyzed in detail, despite both characteristics being relevant, e.g., for sensor choice and placement. This study investigates both characteristics, based on measurements with two earbud models. Spectrally, results indicate that OV-induced earbud vibrations exhibit a low-pass characteristic, with a steep roll-off of \SI{-93}{\deci\bel} per decade above \SI{400}{\hertz}. Thus, sensors with comparatively low noise floors are required to sense the vibrations above \SI{1}{\kilo\hertz}. Spatially, results indicate that the earbuds mainly vibrate in and out of the ear canal entrance, with high consistency between subjects and fits. Simulations confirm that this enables capture of the high-power vibrations below \SI{400}{\hertz} by a single-axis sensor with less than \SI{1.5}{\deci\bel} mean attenuation.
    \end{abstract}
    \vspace{0.5em}

    \begin{center}
    {\small\bfseries Preprint Notice}
    \end{center}
    \begin{quotation}{\small\noindent This PDF is a preprint corresponding to the manuscript originally submitted to the \href{https://aes.org/publications/journal/}{\emph{Journal of the Audio Engineering Society (AES)}} in March 2026. The revised version, following peer review, has been published in \emph{J.~Audio Eng.\ Soc.}, vol.~74, no.~11, pp.~796--810, November 2026, \href{https://doi.org/10.17743/jaes.2026.0301}{doi:10.17743/jaes.2026.0301}. Please cite the published version.}
    \end{quotation}

    \vspace{1cm}

    \end{@twocolumnfalse}
]

\section{Introduction}

Wireless earbuds have become ubiquitous in recent years. Common features like telephony or voice assistants require capturing the wearer's \ac{ov}. However, earbud-mounted microphones are usually affected by acoustic noise and wind. In contrast, \acp{bcm} are robust against said noises \cite{shinSurveySpeechEnhancement2012}, as they capture the \ac{ov} of the wearer as vibrations of the body, i.e. as structure-borne sound. Although the term \textit{body}-conduction is more appropriate in the context of this study, the term \textit{bone}-conduction is used to remain consistent with existing literature.

Utilizing \ac{bc} speech for signal processing applications has been researched extensively. Mainly, \ac{bc} speech has been utilized to support speech enhancement \cite{shinSurveySpeechEnhancement2012,hauretConfigurableEBENExtreme2023a,wangMultimodalSpeechEnhancement2022}. More recently, speech recognition supported by \ac{bc} speech has been researched \cite{heitkaemperBoneConductedSignal2025a}. Other applications include \ac{vad} \cite{schilkInearvoiceMilliwattAudio2023a,heitkaemperBoneConductedSignal2025a}, wind noise reduction \cite{zhouRealTimeDualMicrophoneSpeech2020}, and occlusion effect reduction \cite{weyerFeedbackawareDesignOcclusion2023a}. While traditionally research has focused on picking up \ac{bc} speech using specialized devices \cite{todaStatisticalApproachesEnhancement2012a,zhengNovelThroatMicrophone2018}, e.g., throat microphones, recent research often used the earbud as a pickup location \cite{gaoPracticalEarphoneEavesdropping2023,tagliasacchiSEANetMultimodalSpeech2020d,heBoneconductedVibrationSpeech2023,heitkaemperBoneConductedSignal2025a}. In case of the earbud, \ac{bc} speech is usually captured as vibrations of the earbud housing, using a high-bandwidth accelerometer or \ac{imu}. As earbuds are already widely in use and less cumbersome than specialized devices, they are an attractive pickup location. However, the fundamental characteristics of \ac{ov}-induced earbud vibrations have not yet been widely analyzed.

Important characteristics of the vibration include spectral characteristics, such as the bandwidth of the vibration, as well as spatial characteristics, such as the orientation in space and the variation of said orientation across wearers. Both, spectral and spatial characteristics are practically relevant in multiple contexts: Firstly, they are relevant for the design of the earbud, affecting the choice of an accelerometer, i.e., they influence at which bandwidth and in how many axes the accelerometer should record the vibration, but also in which orientation the accelerometer should be mounted inside the earbud housing. This is especially important as choosing such components entails trade-offs between component cost, processing complexity, \ac{snr}, etc. For example, usually single-axis accelerometers have a better \ac{snr} than multi-axis accelerometers with similar features. In turn, the aforementioned points are relevant for manufacturers of accelerometers. Secondly, spatial and spectral characteristics are relevant for the design of testing equipment, as some modern dummy heads attempt to accurately reproduce the \ac{ov}-induced earbud vibrations, e.g., \cite{headacousticsgmbhProductPageVibridge}. Lastly, these characteristics are important for algorithm design as they inform which added-value the modality of \ac{bc} speech captured at the earbud might provide.

Research exists on different pickup locations \cite{mcbrideEffectBoneConduction2011} as well as on sound transmission in the head in general \cite{stenfeltTransmissionPropertiesBone2005,stenfeltAcousticPhysiologicAspects2011}. One publication models the earbud as a simple mass-spring-system but does not consider spatial orientation or \ac{ov}-induced vibrations \cite{tikanderModelingAttenuationLooselyfit2007a}. Besides the lack of fundamental research, existing publications using earbuds for speech enhancement differ in the way they utilize the accelerometer signal, i.e., by choosing just one axis as input, using all three, or using the norm of all axes \cite{tagliasacchiSEANetMultimodalSpeech2020d,gaoPracticalEarphoneEavesdropping2023,heBoneconductedVibrationSpeech2023}. Accelerometers designed to sense \ac{bc} speech also differ in their supported bandwidth and the number of axes recorded \cite{LIS25BADatasheet2019,V2S200DDatasheet2024}. This reveals a lack of consensus in how the added measurement modality can be utilized in practice. 

A previous pilot study \cite{weyerAnalysisEarbudMountedBoneConduction2024} by the authors provided evidence that the \ac{bc} speech recorded at the earbud exhibits a low-pass characteristic with small device-dependent differences in the cutoff frequency, and that every earbud mostly vibrates in one direction. Although the previous publication by the authors considered five earbud models, the study was limited to a low number of test subjects. This publication extends and corroborates the previous study in the following ways: First, this study focuses on two of the five earbud models, which facilitates to increase the number of participants to support the prior results. Secondly, this study combines the analysis of the spectral characteristics with an analysis of the resulting typical \ac{snr} to assess the most relevant frequency regions. Thirdly, this study employs a measurement procedure to estimate the orientation of the earbuds relative to the wearer's head. This allows to visualize the vibration for all subjects in a joint head-related coordinate system, and thus, to gain a better understanding of the vibration. Lastly, the measured data is used to simulate the loss in signal power introduced by a single-axis accelerometer compared to a three-axis one, which helps to assess the suitability of a single-axis sensor for capturing \ac{ov}-induced vibrations.

This paper is structured as follows: Sec.~\ref{sec:experimentSetup} starts by outlining the experiment setup. Sec.~\ref{sec:orientation} describes how the earbud orientation is estimated. While Sec.~\ref{sec:snr} and Sec.~\ref{sec:sensitivity} examines the spectral characteristics of the vibration, Sec.~\ref{sec:spatial_characteristics} examines the spatial characteristics. Sec.~\ref{sec:projection} investigates the question of the effectiveness of using a single-axis sensor. Sec.~\ref{sec:conclusion} concludes the paper.

\section{Experiment Setup} \label{sec:experimentSetup}

\begin{figure}[tb]
    \centering
    \includegraphics{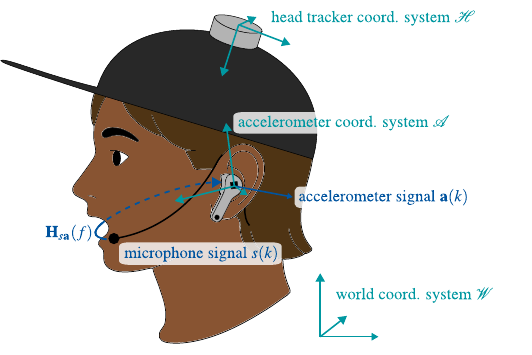}
    \caption{Sketch of the measurement setup, signals and coordinate systems.}
    \label{fig:sketch_measurement_setup}
\end{figure}

The goal of this study is to analyze \ac{ov}-induced earbud vibrations. These vibrations are captured as accelerations of the earbud housing. Note that in the context of this study, the term vibration always refers to the surface acceleration as captured by an accelerometer, and not to the displacement. Thus, earbuds were equipped with accelerometers and measurements of \ac{ov}-induced accelerations were conducted with multiple test subjects. Additionally, a headset microphone was used to capture sound pressure at the mouth as reference. Furthermore, a head tracker was used to capture the orientation of the head. The orientation of the head was captured to estimate the orientation of the earbuds in space and to subsequently express the earbud accelerations in a joint head-related coordinate system. An overview of the measurement setup used is shown in Fig.~\ref{fig:sketch_measurement_setup}. This section defines the recorded signals, and describes utilized hardware, as well as measurement setup and procedure. It concludes with a note on the frequency range considered in this paper for analysis.

\subsection{Signals and Coordinate Systems}
 
This section outlines the signals recorded in this study, which are also shown in Fig.~\ref{fig:sketch_measurement_setup}. The accelerometer signal 
\begin{equation}
    \accelSig=\accelSigVec\in\mathbb{R}^3
\end{equation} 
captures the vibration of the earbud along the three spatial axes of the local coordinate system $\sysAccel$ of the recording accelerometer. Here, $k\in\mathbb{Z}$ is the discrete-time index. The rotation matrix $\earbudOrientationWorld\in\mathbb{R}^{3\times3}$ denotes the orientation of the accelerometer coordinate system $\sysAccel$ in the world coordinate system $\sysWorld$. Furthermore, the rotation matrix $\headtrackerOrientation(l)\in\mathbb{R}^{3\times3}$ captures the orientation of the head tracker coordinate system $\sysHead$ in the world coordinate system $\sysWorld$. Here, $l\in\mathbb{Z}$ is another discrete-time index, usually with much lower sampling rate. Lastly, the headset microphone signal $\headsetSig\in\mathbb{R}$ captures the sound pressure next to the mouth.

\subsection{Hardware}

\begin{figure}[tb]
    \centering
    \includegraphics{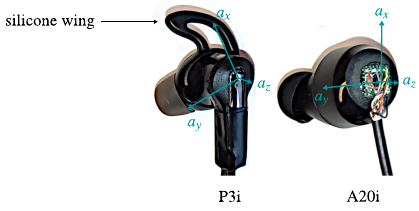}
    \caption{Earbuds used in this study with approximate orientation of accelerometer coordinate system $\sysAccel$.}
    \label{fig:earbuds}
\end{figure}

To capture the \ac{ov}-induced earbud vibrations, one pair of Anker P3i and one pair of Anker A20i earbuds were equipped with three-axis accelerometers. The left earbud of each pair is shown in Fig.~\ref{fig:earbuds}. The earbud models were chosen to represent the housing shapes most common for wireless in-ear headphones on the market today. More specifically, the P3i features a shape with a stem, while the A20i features a knob-like shape. The P3i earbuds were equipped with a silicone \textit{wing} securing them in the concha. This was done as the previous study by the authors showed the best transmission of wearer-induced-vibrations to the earbud for this configuration, as well as the practical consideration that this ensured a secure fit of the earbud with all participants \cite{weyerAnalysisEarbudMountedBoneConduction2024}. In contrast, the A20i was not equipped with a silicone wing, as this is not a feature present for most earbud models.

The ST LIS25BA accelerometer mounted on a custom, minimally-sized PCB was used as accelerometer. The LIS25BA accelerometer is designed to capture \ac{ov}-induced earbud vibrations \cite{LIS25BAApplicationNote2020}. It features a flat bandwidth of up to \SI{2.4}{\kilo\hertz} and records the acceleration in three spatial axes. The accelerometer was mounted inside the P3i earbud, while for the A20i, it was mounted on the outside. The mounting position and orientation is apparent in Fig.~\ref{fig:earbuds} for the A20i, and similar for the P3i. The resulting approximate orientation of the accelerometer coordinate system $\sysAccel$ relative to the earbud is illustrated as well. Note that the chosen accelerometer position corresponds to one which suggests itself to earbud manufacturers. Earbuds commonly feature a main PCB on the inside of the outer surface, handling audio processing, bluetooth, etc. This PCB would be the most readily available mounting position for an accelerometer, avoiding additional cabling inside the earbud, which might be more expensive to manufacture. While smaller, and for the A20i, mounted outside, the PCB used in this study has a similar position and orientation as this main PCB, thus being close to the most relevant point to capture the vibration.

As reference microphone, a Knowles FG-23329-P142 capsule was used. The microphone was attached to a headset arm and positioned next to the subject's mouth. Accelerometers and headset microphone were connected via cables to an audio codec, which in turn was connected via a sound card to a computer. As head tracker, a HTC Vive Tracker 1.0 attached to a baseball cap was used.

\subsection{Measurement Procedure}

Measurements were carried out with 17 subjects and the two pairs of earbuds described in the previous section. Of the 17 subjects, 14 identified as male and 3 as female. The subjects were between 23 and 61 years old, and mostly had a European background. For each combination of subject and earbud, the measurement procedure was repeated once, refitting the earbuds in between. It was aimed to measure realistic fits of the earbuds in the ear canal by instructing the users to position the earbuds in a way that the ear canal is sealed tightly and the earbud's fit is stable. In addition to the earbuds, the subjects were equipped with the headset microphone and the head tracker.

For every fit of a pair of earbuds, three recordings were performed. Firstly, a \SI{5}{\second} long \textit{calibration} recording was performed. The subject was instructed to align their head with the HTC world coordinate system and remain silent and fixed in this position for the duration of the recording. The calibration recording was performed to capture the noise-floor of the worn earbuds and to facilitate the estimation of the earbud orientation. Secondly, a \textit{text} recording was performed. During this recording, the subject read a section of the rainbow passage aloud \cite{fairbanksVoiceArticulationDrillbook1960}. The resulting recordings are \SI{37}{\second} long on average. The subjects were not instructed to speak aloud in any special way, e.g., especially loudly etc. Thirdly, a \SI{5}{\second} long \textit{tilt} recording was performed. The subject was instructed to tilt their upper body and head forward by approximately \SI{30}{\degree}, while not changing the fit of the earbud. They were further instructed to remain silent and fixed in this position for the duration of the recording. The tilt recording was also performed to facilitate the estimation of the earbud orientation.

During all recordings, the headset microphone signal $\headsetSig$ as well as the accelerometer signals $\accelSig$ for the left and right earbud were captured jointly at \SI{48}{\kilo\hertz} sampling rate. Note that left and right earbud are hereafter referred to as left and right side. Combining both sides, this results in a total of $68$ accelerometer signals $\accelSig$ per earbud model. Furthermore, during all three recordings, the orientation of the head was captured as the rotation matrix $\headtrackerOrientation(l)$ with \SI{2}{\hertz} sampling rate.

\subsection{Considered Bandwidth}

The fundamental frequency of human speech is usually above \SI{50}{\hertz} \cite[p. 9]{varyDigitalSpeechTransmission2006}. The minimal fundamental frequency observed in this study was around \SI{100}{\hertz}. Also, below \SI{100}{\hertz}, an earbud mounted accelerometer captures more and more motion artifacts towards lower frequencies. Furthermore, the bandwidth of the ST LIS25BA accelerometer, which has its \SI{-3}{\deci\bel} point approximately at \SI{2.4}{\kilo\hertz}, limits the bandwidth considered in this study. However, higher frequencies are unlikely to be practically relevant as tissue acts as a low-pass due to its damping properties, and air-conducted interference tends to be relatively stronger at higher frequencies, \cite{weyerAnalysisEarbudMountedBoneConduction2024}. The authors thus chose to present results in the range from \SI{75}{\hertz} to \SI{2}{\kilo\hertz}.

\section{Estimation of Earbud Orientation} \label{sec:orientation}

This section outlines how the orientation of the earbuds is estimated. This is done to subsequently express the recorded acceleration in a fixed head-related coordinate system. The coordinate systems discussed here are illustrated in Fig.~\ref{fig:sketch_measurement_setup}.

The accelerometer signal $\accelSig$ is defined in the local coordinate system $\sysAccel$ of the recording accelerometer. For different subjects and different fits of the earbuds, the orientation of this coordinate system $\sysAccel$ relative to the head changes. Furthermore, for different earbuds, there might be systematic differences between the orientation of the earbud inside the concha. For visualization relative to the head and comparison between different earbuds and subjects, it is advantageous to transform the acceleration to a common coordinate system, which is aligned with the head. The world coordinate system $\sysWorld$ of the HTC Vive is used for this purpose. During measurements, the subjects were positioned so that their heads are aligned  with the world coordinate system. To transform the accelerations to this common world coordinate system, an estimate of the orientation $\earbudOrientationWorld$ of the accelerometers, and thus the earbuds, in the world coordinate system is required. This is done using the calibration and tilt recordings as well as the head tracker.

The earbud orientation $\earbudOrientationWorld$ can be estimated by utilizing the fact that the accelerometers also record gravity. The orientation of the gravity vector as it appears in the accelerometer coordinate system for two orientations of the head is recorded. Under the assumptions that a) the orientation of the gravity vector in the world coordinate system is known, b) the orientation of the earbud relative to the head stays constant for both recordings, and c) the rotation between both positions of the head is known, the orientation of the earbud can be estimated. Finding the orientation of the accelerometer in the reference system corresponds to solving Wahba's problem \cite{wahbaLeastSquaresEstimate1965a}. The Kabsch algorithm \cite{kabschDiscussionSolutionBest1978}, which is readily available in standard toolboxes, is used to solve Wahba's problem and calculate $\earbudOrientationWorld$.

The underlying math is briefly outlined here: During the calibration recording, the gravity vector $\gravWorld\in\mathbb{R}^3$ in world coordinates is mapped to accelerometer coordinates via
\begin{equation} \label{eq:grav_one}
    \gravAccelOne = (\earbudOrientationWorldOne)\transposed \gravWorld \in\mathbb{R}^3 .
\end{equation}
Here, the rotation matrix $\earbudOrientationWorldOne\in\mathbb{R}^{3\times3}$ denotes the orientation of the recording accelerometer in the world coordinate system during the calibration recording. Now, the subject tilts their upper body and head forward, so the tilt recording can be performed. Let $\deltaRot\in\mathbb{R}^{3\times3}$ denote the rotation of the head between calibration and tilt recording. As the relative orientation between head and earbuds is fixed, the orientation of the accelerometer in world coordinates during the tilt recording can be written as $\deltaRot \, \earbudOrientationWorldOne$. Thus, during the tilt recording, the gravity vector is mapped to accelerometer coordinates via
\begin{equation} \label{eq:grav_two}
    \gravAccelTwo = (\deltaRot \ \earbudOrientationWorldOne)\transposed \gravWorld = (\earbudOrientationWorldOne)\transposed \deltaRot\transposed \gravWorld \in\mathbb{R}^3 .
\end{equation}
(\ref{eq:grav_one}) and (\ref{eq:grav_two}) form a system of equations centered around the earbud orientation $\earbudOrientationWorldOne$ during the calibration recording. Using the aforementioned Kabsch algorithm, $\earbudOrientationWorldOne$ can be estimated using the paired sets $\{\gravAccelOne,\gravAccelTwo\}$ and $\{\gravWorld,\deltaRot\transposed\gravWorld\}$.

The gravity vectors $\gravAccelOne$ and $\gravAccelTwo$ in the accelerometer coordinate system are obtained as DC components of the calibration and the tilt recording, respectively. The gravity vector in the world coordinate system is known a-priori. The rotation matrix $\deltaRot$ is obtained as
\begin{equation} \label{eq:delta_rot}
    \deltaRot = \headtrackerOrientationTwo (\headtrackerOrientationOne)^{-1} ,
\end{equation}
where $\headtrackerOrientationOne$ and $\headtrackerOrientationTwo$ are the rotation matrices recorded by the head tracker during calibration and tilt measurement, respectively, averaged across the whole length of the corresponding recordings.

\section{Typical Spectra of Earbud Vibration} \label{sec:snr}

The practically useful bandwidth of the accelerometer signal captured at the earbud is commonly limited by its \acl{snr}. In speech enhancement for example, frequency components of the accelerometer signal with $\snr<\SI{0}{\deci\bel}$ are unlikely to provide any benefit. This section investigates \acp{psd} of \ac{ov} and noise components, as well as the resulting \ac{snr}. The \acp{psd} are considered as they enable extending the results to other accelerometers. The goals of this section are: firstly, to identify the frequency range most relevant for applications given the \ac{snr} achievable with state-of-the-art accelerometers, secondly, to derive noise floor requirements at different frequencies, and thirdly, to assess the \ac{snr} resulting from the employed measurement setup.

The analysis of the spatial characteristics of the earbud acceleration requires a three-axis accelerometer, like the ST LIS25BA used in this study. To the best of the authors' knowledge, it has a state-of-the-art noise floor for a three-axis accelerometer suitable to sense \ac{ov}-induced earbud vibrations. However, suitable single-axis accelerometers with lower noise floors exist, and need to be considered to assess achievable \ac{snr}. Thus, two such accelerometers, the Syntiant V2S and the Sonion VPU, are also considered in this section. Their noise floors are derived from publicly accessible documentation \cite{V2S200DDatasheet2024,clemensSonionVoicePick2018}.

\subsection{Methodology and Procedure}

The accelerometer signal $\accelSigIth$ along axis $\accelAxisIdx\in\{x,y,z\}$ is modeled as the sum of \ac{ov}-induced components $\accelSigIthSignal$ and noise $\accelSigIthNoise$. The \acp{psd} of both are denoted by $\psAccelIthSignal$ and $\psAccelIthNoise$, respectively. The noise is defined as all components unrelated to speech production, such as sensor self-noise, and is thus assumed to be uncorrelated with the \ac{ov} component. The \ac{ov} \ac{psd} is estimated as $\psAccelIthSignal=\psAccelIth - \psAccelIthNoise$, where $\psAccelIth$ is obtained from the text recording and $\psAccelIthNoise$ from the calibration recording, during which the subject remained silent. Note that the calibration recording mostly reflects the accelerometer's self-noise floor for frequencies above \SI{100}{\hertz}. Thus, comparison with the self-noise floors of other accelerometers is valid. Values $\psAccelIthSignal<0$ are set to zero, as they are by definition estimation errors. The resulting \ac{snr} can be calculated as $\snr=\psAccelIthSignal/\psAccelIthNoise$. The required \acp{psd} $\psAccelIthSignalPlusNoise$ and $\psAccelIthNoise$ are estimated for every signal $\accelSigIth$, i.e., for every axis and side, and for every text and calibration recording, respectively. The estimation is done using the Welch method \cite[p. 147]{varyDigitalSpeechTransmission2006}, with a $4096$ sample long Hann window, corresponding to around \SI{85}{\milli\second}, and \SI{50}{\percent} overlap. Note that the term power spectral \textit{density} is used in this section to denote normalized long-term average power per unit bandwidth obtained via the Welch method. The normalization is needed to make the power spectrum independent of the analysis window length, facilitating comparison. For visualization, all \acp{psd} are converted to \acp{asd} $\sqrt{\Phi(f)}$, smoothed in $1/12$ octave bands, and displayed in units of \si{\micro\grav/\sqrt{\hertz}}, where $\si{\grav}=\SI{9.81}{\meter/\second^2}$ denotes the acceleration resulting from gravity. The chosen unit facilitates comparison with noise floors of other accelerometers.

\subsection{Results}

\begin{figure}[tb]
    \includegraphics{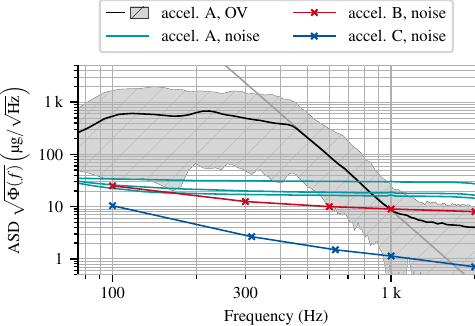}
    \caption{\ac{ov} \acp{asd} $\asAccelIthSignal$ for accelerometer~A. Gray line indicates \SI{-93}{\deci\bel}/decade slope. Noise \acp{asd}}$\asAccelIthNoise$ 
    \label{fig:snr}
\end{figure}

Fig.~\ref{fig:snr} visualizes the \ac{ov} \acp{asd} $\asAccelIthSignal$ and the noise \acp{asd} $\asAccelIthNoise$ computed based on the measurements conducted with the ST LIS25BA accelerometer, labeled accelerometer~A. The \ac{ov} \acp{asd} are displayed as hatched area, encompassing the 1.25th to 98.75th percentile range, and as mean line. Both percentile range and mean are computed across all recordings, sides, and spatial axes. The percentile range is chosen to avoid misrepresentation due to outliers. The mean across axes, recordings, and sides is shown to aid readability. Note that axes- and earbud-specific characteristics are considered in later sections. The noise \acp{asd} are shown as mean per spatial axis, i.e., averaged across recordings and side, as the noise floor differs among axes due to properties of accelerometer~A. Furthermore, the figure shows noise floor \acp{asd} $\asAccelIthNoise$ of the two exemplary single-axis accelerometers, the Syntiant V2S and the Sonion VPU, labeled accelerometer~B and~C, respectively.

The \ac{ov} \acp{asd} $\asAccelIthSignal$ exhibit a low-pass characteristic, with a roughly constant amplitude of around \SI{530}{\micro\grav/\sqrt{\hertz}} between \SI{100}{\hertz} and \SI{400}{\hertz}. The \ac{asd} varies due to differences between axes, earbuds, and loudness of the subjects' voices. Furthermore, the harmonic nature of speech can result in low \ac{asd} values between harmonics, and especially below the fundamental frequency. Above \SI{400}{\hertz}, the \acp{asd} roll off steeply at around \SI{-93}{\deci\bel} per decade. This reflects decreasing speech excitation, but also damping during transmission of the vibration to the earbud. Above \SI{1}{\kilo\hertz}, the roll-off flattens out at mean \ac{asd} of around \SI{5}{\micro\grav}, while the variation increases. Both are likely a result of the estimation error increasing when $\psAccelIthSignalPlusNoise\approx\psAccelIthNoise$. This results in an upward bias of the estimated \ac{ov} \ac{asd} on the \si{\deci\bel} scale, which thus constitutes an upper bound on the true value.

It can be concluded that due to the steep roll-off of the \ac{ov} components, accelerometers with comparatively low noise floors are required to capture components above \SI{1}{\kilo\hertz}. To capture components corresponding to the average \ac{asd} at \SI{20}{\deci\bel}, the noise floor would need to be around \SI{50}{\micro\grav/\sqrt{\hertz}} between \SI{100}{\hertz} and \SI{400}{\hertz}, but only around \SI{0.5}{\micro\grav/\sqrt{\hertz}} between \SI{1}{\kilo\hertz} and \SI{2}{\kilo\hertz}.

The mean noise \acp{asd} $\asAccelIthNoise$ for accelerometer~A are approximately flat between \SI{100}{\hertz} and \SI{2}{\kilo\hertz}, with mean values of around \SI{17}{\micro\grav/\sqrt{\hertz}} for the $x$ and $y$ axis, and around \SI{30}{\micro\grav/\sqrt{\hertz}} for the $z$ axis. The flatness indicates that the \acp{asd} mostly reflect accelerometer self-noise. The \acp{asd} $\asAccelIthSignal$ of the \ac{ov}-induced accelerations fade into the noise floor of the recordings between \SI{500}{\hertz} and \SI{1}{\kilo\hertz}, with the exact crossover depending on recording, side, and axis. The noise \acp{asd}
$\asAccelIthNoise$ for the two exemplary accelerometers B~and~C are lower than for accelerometer A, and also decrease towards higher frequencies while A has a flat characteristic.

\begin{figure}[tb]
    \includegraphics{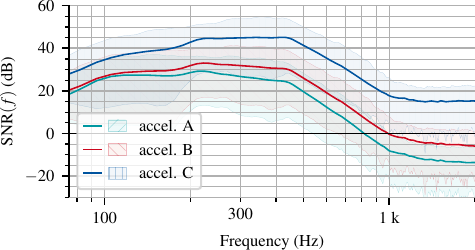}
    \caption{\acp{snr} of \ac{ov} components for accelerometers A--C.}
    \label{fig:real_snr}
\end{figure}

For easier comparison, Fig.~\ref{fig:real_snr} illustrates the resulting \acp{snr} of the \ac{ov} components if they were sensed by the accelerometers A--C. The \ac{snr} is computed by relating the mean and percentile range of the \ac{ov} \ac{asd} shown in Fig.~\ref{fig:snr} to the mean noise \ac{asd} averaged across axes for accelerometer A, as well as to the noise \acp{asd} for accelerometer B~and~C. It can be observed that the \ac{snr} corresponding to the mean \ac{ov} \ac{asd} for accelerometer A~and~B is below \SI{0}{\deci\bel} above \SI{1}{\kilo\hertz}. For accelerometer C, it is positive at around \SI{15}{\deci\bel}. This suggests that an accelerometer with a noise floor level as low as that of accelerometer C might be able to sense \ac{ov}-induced earbud vibrations at $\snr>\SI{0}{\deci\bel}$ up to \SI{2}{\kilo\hertz}. However, whether this is possible in practice cannot conclusively be assessed. The actual \ac{ov} \ac{asd} might be lower due to the upward bias introduced by the estimation, or up to \SI{5}{\deci\bel} higher for optimal alignment of the sensor with the vibration direction. Also, using a single-axis sensor also relies on the vibration direction being constant across frequencies, subjects, and fits. Later sections tackle these questions. It can be concluded that sensing components up to \SI{2}{\kilo\hertz} likely requires a noise floor at least on par with that of accelerometer~C.

Overall, the results indicate: Firstly, high-power components are concentrated between \SI{100}{\hertz} and \SI{400}{\hertz}, and are thus the most application-relevant ones. This is in line with \cite{tagliasacchiSEANetMultimodalSpeech2020d}, where components below \SI{400}{\hertz} provided the largest benefit for speech enhancement. Secondly, a substantially lower noise floor is required to capture components above \SI{1}{\kilo\hertz}. Thirdly, the \ac{snr} limits the bandwidth in which results of later sections are valid to below \SI{1}{\kilo\hertz}.

\section{Bandwidth for Different Earbuds} \label{sec:sensitivity}

The previous section investigated the \textit{absolute} acceleration power which can be captured at the earbud. This section investigates this acceleration \textit{relative} to sound pressure captured using a headset microphone. This way, it is assessed if the typical low-pass characteristic reflects the transmission characteristics of vibrations to the earbud. Furthermore, it facilitates comparison of this transmission characteristics between earbuds, i.e., answering the question if \ac{ov}-induced accelerations can be better captured with some earbuds than others.

To do so, the norm $\tfHeadsetToAccelNorm$ of the (relative) \ac{tf} $\tfHeadsetToAccel$ from the headset microphone to the accelerometer axes is examined. The \ac{tf} is also indicated in Fig.~\ref{fig:sketch_measurement_setup}. The norm $\tfHeadsetToAccelNorm=\sum_{\accelAxisIdx\in\{x,y,z\}} \left| \tfHeadsetToAccelIth \right|^2$ is considered, as the goal is assessing the spectral characteristics independent of direction in this section. $\tfHeadsetToAccelNorm$ measures the total acceleration one can capture at the earbud compared to the sound pressure at the headset microphone, thus quantifying the relation between air- and \acl{bc} speech. The \ac{tf} vector is given by
\begin{equation}
\tfHeadsetToAccel=\tfHeadsetToAccelVec \in \mathbb{C}^3 \,
\end{equation} where $\tfHeadsetToAccelIth\in \mathbb{C}$ are the \acp{tf} from the headset microphone to the accelerometer signal along axis $\accelAxisIdx$. These \acp{tf} $\tfHeadsetToAccelIth$ are computed for every text recording and side using the MMSE-optimal estimator \cite{varyDigitalSpeechTransmission2006}
\begin{equation}
\tfHeadsetToAccelIth = \frac{\cpsHeadsetAccelIthSignal}{\psHeadsetSignal} ,
\end{equation} 
where $\cpsHeadsetAccelIthSignal\in \mathbb{C}$ is an estimate of the cross \ac{psd} of the accelerometer signal $\accelSigIth$ and the headset microphone signal $\headsetSig$, and $\psHeadsetSignal\in \mathbb{R}$ is an estimate of the auto \ac{psd} of the headset microphone signal $\headsetSig$. Both \ac{psd} estimates are obtained by applying the Welch method with parameters as described in the previous section. All resulting \acp{tf} are smoothed in $1/12$ octave bands for better readability.

\begin{figure}[tb]
    \includegraphics{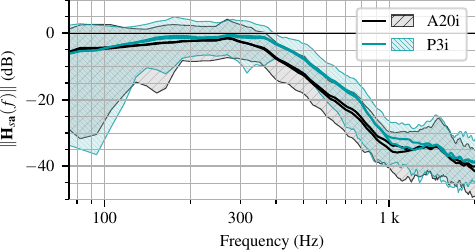}
    \caption{Norm $\tfHeadsetToAccelNorm$ quantifying ratio between total earbud acceleration to sound pressure at the headset microphone.}
    \label{fig:tf_sum_headphones}
\end{figure}

Fig.~\ref{fig:tf_sum_headphones} displays the norm $\tfHeadsetToAccelNorm$ for both earbud models. Lines show the mean norm per earbud model and side, i.e., averaged across subjects and fits. Note that this results in two curves per earbud model, one for the left and one for the right side. Hatched areas indicate the 5th to 95th percentile range per earbud model, i.e., averaged across subjects, fits, and sides. The percentile range is chosen again to avoid misrepresentation due to outliers. First, a similar low-pass characteristic can be observed for all recordings, with limited variation around the mean above \SI{200}{\hertz}. The presence of outliers below \SI{200}{\hertz} is because this is below the fundamental frequency of some subjects' voices. The mean curves show a similar low-pass behavior for both earbud models, with high consistency between left and right side. The most notable difference is the cutoff frequency, which begins around \SI{280}{\hertz} for the A20i, and around \SI{360}{\hertz} for the P3i. Afterwards, the mean lines roll off with around \SI{-58}{\deci\bel}/decade.

This confirms the results from the previous study conducted by the authors \cite{weyerAnalysisEarbudMountedBoneConduction2024}: \ac{bc} speech captured at the earbud exhibits a low-pass characteristic, with a slight earbud-dependent difference in the cutoff frequency. It can also be concluded that the \ac{psd} observed in the previous section is in part a result of transmission properties of the tissue and the earbud, which attenuates components above \SI{400}{\hertz}. The differences in cutoff are likely a result of the mass of the earbud and coupling to the concha and ear canal, with lower mass and better coupling shifting the cutoff to higher frequencies. Here, the P3i likely has better coupling due to the silicone wing securing it in the concha.

\section{Spatial Characteristics of Vibration} \label{sec:spatial_characteristics}

When equipping an earbud with an accelerometer to sense \ac{ov}-induced vibrations, one needs to choose an accelerometer with either one, two, or three axes, and one needs to decide how to position the accelerometer inside the earbud. For both choices, the spatial characteristics of the \ac{ov}-induced vibrations, i.e., how the earbud vibrates in space, is important. Thus, this section investigates these characteristics using a \ac{pca} applied to the text recordings. The \ac{pca} finds the main spatial directions in which the earbud accelerates during speech production. After outlining the applied \ac{pca} procedure, this section evaluates if the earbuds mainly vibrate along one direction in space and if this direction is similar across different subjects and fits. Both analyses serve as a basis for Sec.~\ref{sec:projection}, as capturing the vibration with a single-axis sensor is only possible, if there is a main vibration direction which is sufficiently constant across subjects and fits. In the last part of this section, the main vibration direction is visualized in a joint head-related coordinate system. This is done to compare the vibration direction across earbud models and to give a more intuitive understanding of the vibration direction. 

The choice of coordinate system differs between subsections. For the analysis of the directionality in Sec.~\ref{sec:directionality}, it is irrelevant. The analysis of the variance of the main direction in Sec.~\ref{sec:variance_dir} is performed in the coordinate system $\sysAccel$ of the recording accelerometer. This is done because in practice, the accelerometer signal $\accelSig$ is always recorded in the system $\sysAccel$, thus, direction variances in this system directly result in attenuation when a single-axis sensor is used. For the visualization of the vibration direction in Sec.~\ref{sec:vis_main_direction}, the directions are transformed into the joint head-related coordinate system $\sysWorld$. Firstly, this allows comparing both earbud models and both sides, and secondly, it is more intuitive compared to the accelerometer coordinate system $\sysAccel$, which changes its relative orientation to the head with subject, fit, earbud, and side. 

\subsection{Procedure}

\ac{pca} is performed on every accelerometer signal $\accelSig$ obtained from a text recording as described in Sec.~\ref{sec:experimentSetup}. One three-axis accelerometer signal $\accelSig$ is given for each text recording, i.e.,  for every subject, earbud, fit, and for the left and right side, respectively. Every signal is prefiltered with an FIR bandpass with $4097$ coefficients, spanning \SI{100}{\hertz} to \SI{1.5}{\kilo\hertz}. The lower bound of \SI{100}{\hertz} is chosen as below \SI{100}{\hertz}, there is not much speech excitation to be expected, however, there may be signal due to motion artifacts which would bias the identified direction. The upper bound of \SI{1.5}{\kilo\hertz} is chosen, because beyond \SI{1}{\kilo\hertz} the signal is mainly constituted by accelerometer self-noise, and a slightly larger frequency range is considered to provide some additional margin. For every accelerometer signal $\accelSig$, and thus, for every subject, earbud, fit, and side, three \acp{pc} $\pcIth\in\mathbb{R}^3$ and their corresponding variances $\varIth$ are obtained, with $n\in\{1,2,3\}$. The ratio $\rho_n=\varIth/\sum_{n'=1}^3 \sigma_{\mathbf{a}_{\text{p}n'}}^2$ gives the part of the total variance explained by the $n$th component.

\subsection{Directionality} \label{sec:directionality}

This section investigates if, for every recording, there is a direction in space along which the earbud mostly vibrates. To investigate this, the variances explained by the \ith \ac{pc} are considered. 

The \ac{pca} identifies the direction in space along which the accelerometer signal $\accelSig$ has its biggest variance as first \ac{pc} $\pcFirst$. Now, if most of the signal's variance can be explained using this direction $\pcFirst$, then the earbud mainly accelerates along this direction. If the second \ac{pc} $\pcSecond$ also explains a significant portion of the total variance, then the earbud accelerates along different directions in a plane spanned by the first and second \ac{pc}. 

\begin{table}
    \centering
    \caption{Ratio $\rho_n$ of the variance explained by the \ith \ac{pc} $\pcIth$ in percent, computed per recording and side. Shown are the mean and minimum-to-maximum range per earbud model across recordings and sides. Values are given in the form ``mean (minimum -- maximum)''. }\label{tab:mag_ratio_first}
    \begin{tabular}{ccc}
    \toprule
    component $n$ & A20i & P3i \\
    \midrule
    1 & 79 (49\,--\,92) & 89 (77\,--\,96) \\
    2 & 16 (5\,--\,44)  &  9 (3\,--\,21) \\
    3 &  5 (1\,--\,13)  &  2 (1\,--\,6) \\
    \bottomrule
    \end{tabular}
\end{table}

Tab.~\ref{tab:mag_ratio_first} gives the mean as well as the minimum-to-maximum range of the ratio $\rho_n$ of the variance explained by the \ith \ac{pc} $\pcIth$ in percent. For the P3i, the first \ac{pc} $\pcFirst$ explains \SI{89}{\percent} of the total variance on average. Thus, for most of the recorded signals, most power is indeed oriented along one spatial axis. Furthermore, the second \ac{pc} $\pcSecond$ explains \SI{9}{\percent} and the third \ac{pc} $\pcThird$ explains \SI{2}{\percent} of the total variance on average. Thus, for some recorded signals a second direction is required to describe most of the variance, while the third direction is mostly irrelevant. This corresponds to the earbud accelerating in a plane, however, most acceleration is still oriented along one direction.

For the A20i, the results are overall similar. However, with the first \ac{pc} $\pcFirst$ only explaining \SI{79}{\percent} of the total variance on average, the acceleration is somewhat less oriented along one direction overall. It is noteworthy that the second \ac{pc} $\pcSecond$ explains up to \SI{44}{\percent} of the total variance for some recorded signals. Thus, for some recorded signals, first and second \ac{pc} explain similar amounts of variance. Thus, the earbud accelerates more or less equally into different directions in the plane. 

It can be concluded that there is indeed a direction along which most of the acceleration power is oriented for most accelerometer signals $\accelSig$, i.e., for most combinations of earbud, subject, fit, and side. For some accelerometer signals $\accelSig$, especially for the A20i, a second direction is also relevant, which corresponds to the earbud accelerating along different directions in a plane. This could be caused by the earbud accelerating into different directions at different frequencies, but also by phase shifts between axes at a given frequency, or a combination of both. Note that phase shifts between axes correspond to the acceleration vector rotating in space. Physically, the observed acceleration could well be explained by the earbud vibrating along an elliptical trajectory, which is more elongated for the P3i, and less elongated for the A20i.

\subsection{Variance of Main Direction} \label{sec:variance_dir}

The result of the previous section is that the recorded acceleration is mainly constituted by one directional component. The question arises, how this direction varies between recordings, especially between subjects and fits. This is important as a large between-fit variation would result in a higher attenuation if a single-axis accelerometer is used to sense the vibration. Thus, the angle $\alpha=\arccos{(\pcFirst\transposed\, \pcMeanFirst / (\norm{\pcFirst}\,\norm{\pcMeanFirst}))}$ between individual first \acp{pc} $\pcFirst$ and the mean \ac{pc} direction $\pcMeanFirst$ is considered. The mean direction is obtained by averaging all \acp{pc} $\pcFirst$ per earbud and side, i.e.,  across all subjects and fits, and then normalizing to length one. Note that sign-ambiguity of the \ac{pca} is resolved prior to averaging by sign-flipping all vectors so they are closest to a common mean. Further, note again that the angle $\alpha$ is calculated in the accelerometer coordinate systems $\sysAccel$, as this is what is practically relevant when capturing the acceleration.

With \SI{6.5}{\degree}, the mean of the angular distance $\alpha$ to the mean direction is remarkably small for the P3i, and still only \SI{14.7}{\degree} for the A20i. The lower variation in the direction for the P3i is probably due to the presence of the silicone wing, securing it in the concha and prohibiting larger rotation. The maximum of the angular distance $\alpha$ to the mean direction is \SI{22.2}{\degree} for the P3i and \SI{38.9}{\degree} for the A20i. 

Overall, it is concluded that deviation from the mean direction is small on average. These results are in line with the results of the earlier study conducted by the authors \cite{weyerAnalysisEarbudMountedBoneConduction2024}. Furthermore, even the directions with maximum deviation are within a $\pm\SI{45}{\degree}$ cone, and thus, far from being orthogonal to the mean direction. This indicates that choosing the mean direction as axis along which the acceleration is captured would not introduce large attenuations. Sec.~\ref{sec:projection} investigates this empirically.

\subsection{Visualization of Main Direction} \label{sec:vis_main_direction}

So far, this study has considered acceleration only in the local coordinate system $\sysAccel$ of the recording accelerometer. However, this leaves the question open, how the earbud vibrates relative to the head. Thus, this section visualizes the main directions of acceleration relative to the head, using the orientation of the accelerometers estimated as described in Sec.~\ref{sec:orientation}. The first \acp{pc} $\pcFirstWorld$ in world coordinates are given by $\pcFirstWorld = (\earbudOrientationWorldOne)\transposed \pcFirst$, where the rotation matrix $\earbudOrientationWorldOne$ gives to the orientation of the recording accelerometer during the corresponding calibration recording, where the head was aligned with the world coordinate system $\sysWorld$.

\begin{figure}[tb]
    \centering
    \includegraphics[trim=1.65cm 1cm 0 1cm]{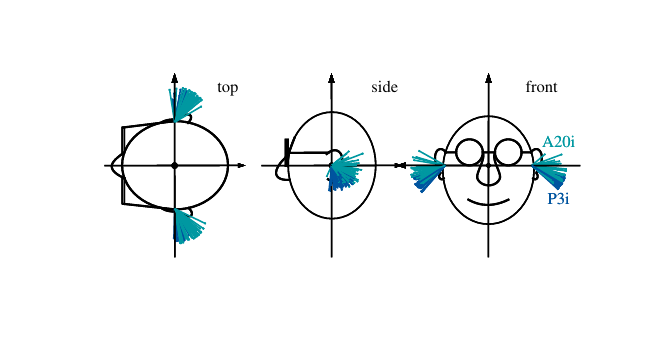}
    \caption{First \acp{pc} $\pcFirstWorld$ in head-aligned coordinate system $\sysWorld$.}
    \label{fig:pcs_world} 
\end{figure}

Fig.~\ref{fig:pcs_world} shows the first \acp{pc} $\pcFirstWorld$ for every text recording and every side. A head is sketched in the background for reference. The \acp{pc} are shifted to the positions of the ears. The head is shown from three points of view. It can be observed that the main direction of acceleration clearly clusters per earbud model and is largely symmetrical between both sides. However, even across earbud models, the direction is similar. The main direction of acceleration is to the sides of the head, with a slight tendency back- and downwards, where the P3i has a tendency downward and the A20i has a tendency backwards. The validity of the results is supported by the fact that the main direction of acceleration is largely symmetric between left and right side, as well as by the fact that it is physically plausible, as it corresponds to a vibration in and out the ear canal entrance. 

Overall, it is observed that the earbud accelerates out of the ear canal to the sides of the head for both earbud models, with a downwards tendency for the P3i, and a backwards tendency for the A20i. Note that as the \ac{pca} is computed based on the time-domain signals, the identified main direction of vibration mostly reflects the high-power components below \SI{400}{\hertz}. The analysis presented in the next section considers frequency-dependent behavior.

\section{Simulation of Single-Axis Accelerometer} \label{sec:projection}

The last section examined the spatial properties of the \ac{ov}-induced acceleration. In contrast, this section examines the more application-oriented question if and under which conditions one can record this vibration effectively with a single-axis accelerometer. In order to analyze this, the projection of the three-axis signal onto different directions is simulated, and the resulting attenuation per frequency assessed.

It is simulated what a single-axis accelerometer would have recorded by calculating the projection 
\begin{equation}
    \projSig = \direction\transposed\cdot\accelSig\in\mathbb{R}
\end{equation}
of the three-axis accelerometer signal onto the direction $\direction\in\mathbb{R}^3$ with $\norm{\direction} = 1$. The attenuation resulting from the projection is assessed via the \ac{psg} 
\begin{equation}
    \pGain=\frac{\psProjected}{ \psAccelSum } \in\mathbb{R},
\end{equation} 
where $\psProjected\in\mathbb{R}$ is the \ac{psd} of the projected signal $\projSig$, and $\psAccelSum=\psAccelSumExpanded$ is the sum of the \acp{psd} $\psAccelIth\in\mathbb{R}$ of the accelerometer signal across the spatial axes. The \acp{psd} are estimated using the Welch method with the parameters as described in Sec.~\ref{sec:snr}. An overall high \ac{psg} $\pGain$ for direction $\direction$ indicates that most of the total vibration power can be captured by a single-axis accelerometer mounted towards direction $\direction$. Reasons for a low \ac{psg} include the direction $\direction$ being orthogonal to the vibration direction, but also the vibration power being distributed along multiple spatial axes. This can be caused by the acceleration vector rotating in space for one frequency, or the acceleration vector at different frequencies pointing into different directions.

\subsection{\ac{pc}-Based Directions} \label{sec:pc-based-directions}

This section investigates how well the directions identified in Sec.~\ref{sec:spatial_characteristics} using \ac{pca} perform as sensor mounting directions. This is done to showcase the attenuation introduced by a favorably-chosen and an unfavorably-chosen mounting direction. Additionally, this also allows to further investigate the spatial characteristics of the vibration in a frequency-dependent way.

The mean directions $\pcMeanFirst$ and $\pcMeanThird$ of the first and third \ac{pc}, respectively, are considered as mounting directions $\direction$. The mean direction of the first \acp{pc} is chosen, because mounting the sensor towards the average main axis of vibration seems most effective for capturing the vibration. The mean direction of the third \ac{pc} is chosen to showcase attenuation introduced by unfavorably-chosen mounting directions. The mean directions are calculated as in the previous section per earbud model and side, i.e., by averaging across all subjects and fits, and renormalizing to unit length afterwards. The \ac{psg} $\pGain$ is calculated by projecting the signals from the text as well as the calibration recordings onto the mean directions derived based on the text recording as explained earlier. Note that the \ac{psg} for projection of the calibration recording is calculated to indicate for which frequencies the \ac{psg} $\pGain$ mostly reflects the noise floor.

\begin{figure}[tb]
    \centering
    \begin{subfigure}{\linewidth}
        \centering
        \includegraphics{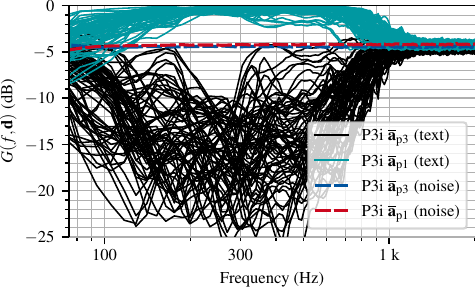}
        \caption{P3i earbud.}
        \label{fig:beamform_gain_P3i}
    \end{subfigure}
    \hfill
    \begin{subfigure}{\linewidth}
        \centering
        \includegraphics{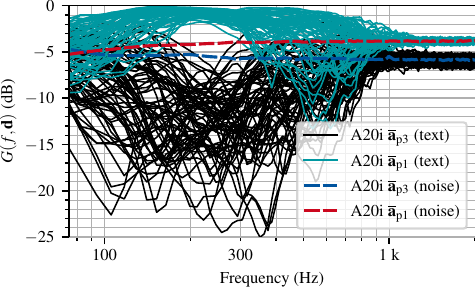}
        \caption{A20i earbud.}
        \label{fig:beamform_gain_A20i}
    \end{subfigure}
    \caption{Attenuation as indicated by the \acp{psg} $\pGain$ resulting from projecting text recordings onto single-axis accelerometer mounted towards mean directions $\pcMeanFirst$ and $\pcMeanThird$. Mean \acp{psg} based on noise floor recordings for comparison. Note that as \acp{psg} are depicted, no \ac{snr} can be inferred by comparing text to noise curves.}
    \label{fig:beamform_gain}
\end{figure}

Fig.~\ref{fig:beamform_gain_P3i} and Fig.~\ref{fig:beamform_gain_A20i} show the \ac{psg} $\pGain$ for the P3i and A20i, respectively. While curves marked with (\textit{text}) refer to the \ac{psg} for an individual text recording, curves marked with (\textit{noise}) refer to the mean \ac{psg} averaged per earbud model across all calibration recordings and both sides. 

For the P3i, the favorably-chosen direction P3i~$\pcMeanFirst$ results in remarkably little attenuation by the projection, with a mean attenuation of \SI{0.7}{\deci\bel} in the range between \SI{100}{\hertz} and \SI{750}{\hertz}. In contrast, the attenuation is much higher for the unfavorably-chosen direction P3i~$\pcMeanThird$, with a mean of \SI{14}{\deci\bel} in the same frequency range. For the favorably-chosen mounting direction, the variance is also rather low in this range. Notably, the variance increases below \SI{200}{\hertz}, towards lower frequencies. Above \SI{600}{\hertz}, the \ac{psg} begins to decay to its noise floor level, which makes it increasingly hard to draw conclusions above that frequency.

For the A20i, the favorably-chosen direction A20i~$\pcMeanFirst$ still results in less attenuation than the unfavorably-chosen A20i~$\pcMeanThird$, with the former having a mean attenuation of \SI{2.4}{\deci\bel} in the range between \SI{100}{\hertz} and \SI{750}{\hertz}, and the latter having a mean attenuation of \SI{9.7}{\deci\bel} in the same frequency range. However, the variation of the attenuation for the favorably-chosen direction is much higher, as is especially visible around \SI{600}{\hertz}, where the projection introduces up to \SI{13}{\deci\bel} of attenuation for some recordings. In contrast, the attenuation around this frequency introduced by the unfavorably-chosen direction A20i~$\pcMeanThird$ actually is often lower than the attenuation introduced by the favorably-chosen direction.

For consistency with the rest of the paper, values are now also reported for the high-power region between \SI{100}{\hertz} and \SI{400}{\hertz}. Here, the favorably-chosen direction yields \SI{0.7}{\deci\bel} and \SI{1.5}{\deci\bel} mean attenuation for the P3i and A20i, respectively, while the unfavorably-chosen direction yields \SI{15}{\deci\bel} and \SI{11}{\deci\bel}. Note that the previously reported range up to \SI{750}{\hertz} was chosen to show that attenuation remains low even above \SI{400}{\hertz} to some extend.

The results suggest that the vibration indeed consists of a main component along a main direction, which can be well captured with a single-axis sensor. For the P3i, this works across a larger frequency range than for the A20i. For the A20i, the favorably-chosen direction introduces attenuations around \SI{600}{\hertz}. This suggests a more complex spatial behavior consistent with the findings in Sec.~\ref{sec:directionality}, where possible interpretations were already discussed. Again, the silicone wing securing the P3i in the concha could be the cause for its more uniform vibration direction. Notably, the attenuation also increases below \SI{200}{\hertz} towards \SI{100}{\hertz} for both earbud models, again suggesting more complex spatial behavior around there as well. It should be noted though, that \SI{100}{\hertz} is also already below the fundamental frequency of the voice of some subjects, which likely contributes to increased variation at this frequency. If one uses a single-axis sensor, choosing the direction based on the first \ac{pc} still seems overall much favorable to choosing the one based on the third \ac{pc}, as the latter attenuates the signal stronger on average. How components towards and above \SI{1}{\kilo\hertz}, which could possibly be sensed with a single-axis accelerometer due to a lower noise floor, are attenuated by projection onto the single axis cannot be answered by this analysis, due to the limitations imposed by the noise of the used three-axis accelerometer.

To make the attenuation resulting from using a single-axis accelerometer more vivid, instead of the \ac{psg}, the \acp{psd} of original and projected signals are now considered for an example subject. As the \ac{psg} is normalized to the total vibration power, it enables comparing attenuation for subjects with different voice spectra. However, the \acp{psd} add more context information, e.g., indicating the \ac{snr} of the projected signal more directly. To select the subject, the mean \ac{psg} $\pGain$  is considered. The mean $\pGain$ is calculated by averaging per subject across all text recordings, both sides, and all frequencies between \SI{100}{\hertz} and \SI{1}{\kilo\hertz}. The subject was selected whose mean $\pGain$ is closest to the mean $\pGain$ averaged across all subjects. 

\begin{figure}[tb]
    \centering
    \includegraphics{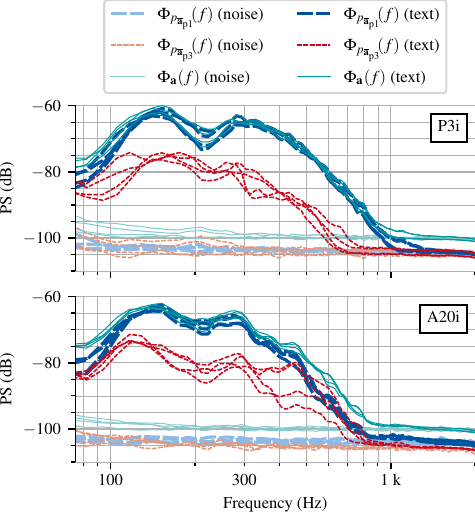}
    \caption{\acp{psd} of simulation of single-axis sensor for an exemplary subject and both earbud models.} \label{fig:beamform_psd_example}
\end{figure}

Fig.~\ref{fig:beamform_psd_example} shows the \ac{psd} $\psAccelSum$, corresponding to the total power summed across all three spatial axes, as well as the \acp{psd} $\psProjPcFirst$ and $\psProjPcThird$ of the accelerometer signal $\accelSig$ projected on the same directions as before. All \acp{psd} are shown for the text recording (\textit{text}), as well as for the calibration recording (\textit{noise}). Note that the peaks and dips observable in the \acp{psd} for the text recordings reflect the typical fundamental frequency of the subject. It can be observed that for the P3i, a near optimal capture is achieved by the single-axis accelerometer, except for lower frequencies. In comparison, the A20i has a suboptimal capture for frequencies above \SI{300}{\hertz}. However, overall the major part of the high-power components is captured by the favorably-chosen directions of every earbud and side. In contrast, the unfavorably-chosen directions do not achieve this goal.

Overall, it is concluded that if a suitable direction is known, a good capture of the high-power components between \SI{100}{\hertz} and \SI{400}{\hertz} is possible with a single-axis accelerometer. However, how much using a single accelerometer attenuates the signal varies with the earbud model. Namely, for the A20i it seems like capturing higher frequencies accurately, a three-axis accelerometer might be needed.

\subsection{Higher-Frequency Attenuation for A20i}

The question arises, how it is possible to capture the higher frequencies for the A20i. This section investigates this question using two approaches: First, it is analyzed if a better accelerometer mounting direction for the A20i can be found, by directly optimizing the broadband sum of the \ac{psg}. This investigates if a direction can be found, which is equally good at sensing most frequency components. Secondly, it is investigated if considering the second or third \ac{pc} in addition to the first one results in substantially less broadband attenuation. This informs how many axes need to be recorded to capture the full bandwidth of the vibration.

First, the \ac{psg}-based optimization is considered. One possible reason for the attenuation of higher frequencies is that the \ac{pca} considers the time-domain acceleration directly. Thus, vibrations at frequencies, where the signal has more power, are weighted higher by the \ac{pca}. To counteract this, this section investigates finding a direction which maximizes the sum of the \ac{psg}. As the \ac{psg} is normalized to the total power, this demonstrates if a direction can be found which captures all frequency components equally well.

\begin{figure}[tb]
    \centering
    \includegraphics{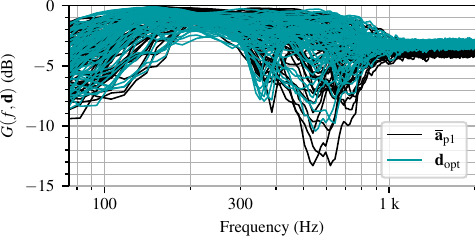}
    \caption{\ac{psg} for broadband-optimal and \ac{pca}-derived projection direction for A20i earbud.}
    \label{fig:projection_pca_vs_opt_A20i}
\end{figure}

In order to do this, the \ac{psg} $\pGain$ is computed for the set $\dSet=\{\direction_i\}$ of $3250$ directions, sampled on a hemisphere. Sampling on a hemisphere is sufficient, as the mathematical projection is by definition symmetric. The directions are sampled using the Fibonacci sphere algorithm, resulting in every point having a maximal angular distance of \SI{2.5}{\degree} to its nearest neighbor on the sphere. Let $\pGainAveraged$ denote the mean \ac{psg} per earbud and side, i.e., averaged across subjects, fits, and a frequency range. The optimal direction $\directionOpt$ for the frequency range between \SI{100}{\hertz} and \SI{1}{\kilo\hertz} on $\dSet$ is found for every side of the A20i as $\directionOpt = \text{arg max}_{\direction\in\dSet} \pGainAveraged$. The resulting \ac{psg} $\pGain$ is shown in Fig.~\ref{fig:projection_pca_vs_opt_A20i} next to the \ac{psg} for the  \ac{pc}-derived direction A20i~$\pcFirst$. No striking differences are observed. Thus, it is concluded that it is not possible to find one spatial direction with lower broadband attenuation for the A20i.

\begin{figure}[tb]
    \centering
    \includegraphics{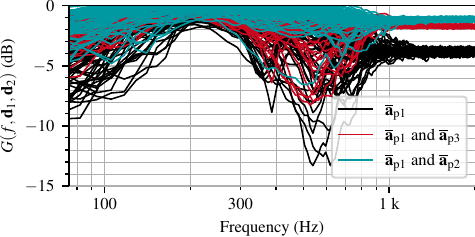}
    \caption{Effect of considering additional \ac{pc} directions on \ac{psg} for A20i earbud.}
    \label{fig:projection_pca_other_dirs_A20i}
\end{figure}

Now, the other \ac{pc} directions are considered. Results from Sec.~\ref{sec:directionality} suggest that a second direction is required to capture most of the acceleration signal's variance. Thus, the question arises if one can reduce the attenuation around \SI{600}{\hertz} specifically by considering a second direction. In order to evaluate this, a two-degree-of-freedom \ac{psg} is considered, given by
\begin{equation}
\pGainModified= \frac{\Phi_{\direction_1}(f) + \Phi_{\direction_2}(f)}{ \psAccelSum } \in\mathbb{R} .
\end{equation}
It quantifies, in the case of orthogonal directions $\direction_1$ and $\direction_2$, how much acceleration power is in the plane spanned by the directions $\direction_1$ and $\direction_2$. The directions $\direction_1=\pcMeanFirst$, and $\direction_2=\pcMeanSecond$ or $\direction_2=\pcMeanThird$ are chosen. This quantifies how much attenuation is introduced if one takes either the second or the third \ac{pc} into account. Fig.~\ref{fig:projection_pca_other_dirs_A20i} shows the resulting two-degree-of-freedom \ac{psg} $\pGainModified$. It can be observed that if the second \ac{pc} is taken into account in addition to the first one, the attenuation around \SI{600}{\hertz} is reduced more for many recordings than if the third \ac{pc} is taken into account. It can be concluded that accurate capture at higher frequencies requires a multi-axis accelerometer in case of the A20i.

\subsection{Deviation from Optimal Direction}

The results from Sec.~\ref{sec:pc-based-directions} suggested that a favorably-chosen direction enables capturing a substantial amount of the high-power components. The question arises how to identify this direction in practice when designing an earbud. One possibility would be making measurements with a three-axis accelerometer attached to the housing of the earbud in question. If that is too cumbersome, Sec.~\ref{sec:spatial_characteristics} offers information on an a-priori guess for a good direction, as it deals with the spatial characteristics of the \ac{ov}-induced earbud vibration. Also, practical constraints might play a role in choosing the mounting direction, i.e., mounting on the main PCB might be favorable production-wise. The follow-up question arises of how a deviation from the optimal mounting direction affects the amount of acceleration power recorded by a single-axis sensor, e.g., if small deviations already result in strong attenuations. This section investigates this by assessing the spatial distribution of the \ac{psg}.

To assess the spatial distribution, the mean \ac{psg} $\pGainAveraged$ is considered again. It is again averaged across subjects and fits, but this time between \SI{100}{\hertz} and \SI{400}{\hertz} to capture the attenuation of the high-power components. The mean \ac{psg} is calculated for all directions in the set $\dSet$, which has been introduced in the previous section. To obtain values on the full sphere, the computed mean \acp{psg} $\pGainAveraged$ are mirrored at the origin. For visualization, the resulting sphere is rotated by the mean orientation $\earbudOrientationAveraged$, so that the result is expressed relative to the head. Note that in contrast to Sec.~\ref{sec:spatial_characteristics}, the individual orientations $\earbudOrientationWorld$ per recording are not used for the transformation as this would change the direction variance of $\pGainAveraged$. This is not desired here, as in the application, the variance in the accelerometer coordinate system, where the actual signals are observed, is relevant. To assess the attenuation relative to the optimal mounting direction, the \ac{psg} $\pGainAveraged$ is normalized to its maximum  $\pGainAveragedLabel(\directionMax) = \max_{\direction\in\dSet} \pGainAveraged$, which it obtains for the optimal direction $\directionMax$, resulting in the normalized mean \ac{psg} $\pGainAveragedNormalized$.

\begin{figure}[tb]
    \begin{subfigure}{0.225\linewidth}
        \centering
        \includegraphics[trim=0.25cm 0 0.25cm 0.1cm,clip]{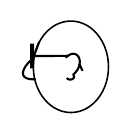}
        \caption{Head}
        \label{fig:projection_polar_head}
    \end{subfigure}
    \begin{subfigure}{0.3\linewidth}
        \centering
        \includegraphics[trim=0 0 1.7cm 0,clip]{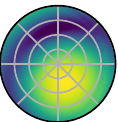}

        \caption{P3i}
        \label{fig:projection_polar_P3i}
    \end{subfigure}
    \begin{subfigure}{0.35\linewidth}
        \includegraphics{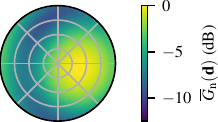}
        \caption{A20i}
        \label{fig:projection_polar_A20i}
    \end{subfigure}
    \caption{Spatial distribution of normalized mean \ac{psg} $\pGainAveragedNormalized$ on left hemisphere, with sketch of head for reference.}
    \label{fig:projection_polar}
\end{figure}

Fig.~\ref{fig:projection_polar} shows the normalized mean \ac{psg} $\pGainAveragedNormalized$ of the left P3i and A20i earbud, respectively. The sketch of the head shown in Fig.~\ref{fig:projection_polar_head} acts as a reference; the view on the polar plot is the same as on the head, i.e., the reader views the left hemisphere from the right, where the center of the polar plot corresponds to the axis connecting both ears through the head. It can be observed that there is a large area in which the gain is close to \SI{0}{\deci\bel} for both earbud models. A band of low average gain is also observed. This band corresponds to the circle of directions perpendicular to the optimal direction. For the P3i, the gain is minimally \SI{-12}{\deci\bel}, and for the A20i \SI{-8.8}{\deci\bel}. The lower attenuation for the A20i is likely a result of the higher variation of the vibration direction.

The previous paragraph showed the full spatial distribution of the normalized mean \ac{psg} $\pGainAveragedNormalized$. For practical purposes, it is helpful to tie the expected attenuation to an angle, i.e., to assess which angular distance $\alpha$ to the optimal mounting direction results in how much additional attenuation. Recall that the direction $\directionMax$ corresponds to the mounting direction of a single axis sensor which results in the least attenuation. For a specific $\alpha$, all directions $\direction$ having an approximate angular distance $\alpha$ to the maximum direction $\directionMax$ lie on contour rings on the sphere. All directions $\direction$ lying on a ring with angular width $2\epsilon$ and mean angular distance $\alpha$ to the maximum direction $\directionMax$ are given by the set
\begin{equation}
    \dSetAlpha = \{\direction \in \dSet \mid \left| \arccos(\direction, \directionMax) - \alpha \right| < \epsilon\}.
\end{equation}
Using this, the value range of the normalized mean \ac{psg} on a contour ring with mean angular distance $\alpha$ to the optimal direction $\directionMax$ and angular width $2\epsilon$ is given by the minimum normalized mean \ac{psg}
\begin{equation}
    \pGainAveragedMinAlpha = \min_{\direction\in\dSetAlpha} \pGainAveragedNormalized
\end{equation}
and the analogously defined maximum normalized mean \ac{psg} $\pGainAveragedMaxAlpha$. This quantifies how much attenuation one encounters minimally and maximally, when one mounts the accelerometer with an angular distance $\alpha$ from the optimal direction $\directionMax$. Here, $\alpha$ is sampled in \SI{2.5}{\degree} steps, and $\epsilon=\SI{1.25}{\degree}$ is chosen accordingly.

\begin{figure}[tb]
    \centering
    \includegraphics{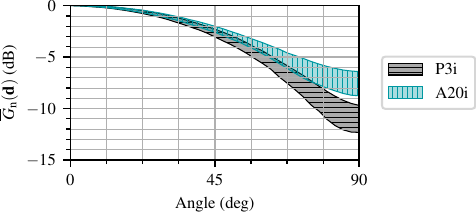}
    \caption{Minimum-to-maximum range of normalized mean \ac{psg} $\pGainAveragedNormalized$ on contour rings $\dSetAlpha$ with angular distance $\alpha$ to optimal direction $\directionMax$.}
    \label{fig:projection_polar_linecut_P3i}
\end{figure}

Fig.~\ref{fig:projection_polar_linecut_P3i} shows the minimum to maximum range of the normalized mean \ac{psg} $\pGainAveragedNormalized$, depending on the angular distance $\alpha$ to the optimal direction $\directionMax$. To simplify the visualization, the maxima and minima across left and right side are shown. Note that only angular distances $\SI{0}{\degree}\leq\alpha\leq\SI{90}{\degree}$ are considered due to symmetry. It can be observed that the normalized mean \ac{psg} is minimally \SI{-3}{\deci\bel} for an angular distance $\alpha$ of up to $\pm\SI{45}{\degree}$ to the optimal direction for the P3i, and \SI{-2.7}{\deci\bel} for the A20i. 

The results suggest that while it is important to approximately choose the correct mounting direction, there is a large margin for error for the considered earbud models. Only large angular distances to the optimal direction, which correspond to the sensing axis being perpendicular to the axis of main vibration, result in large attenuations of up to \SI{12}{\deci\bel} on average. The result is mathematically intuitive: a single-axis sensor measures the projection of the vibration onto its sensing direction. In the idealized case of a pure vibration along a straight line in space for one frequency, the attenuation at that frequency follows $\cos(\alpha)$, which results in \SI{3}{\deci\bel} attenuation at \SI{45}{\degree} and infinite attenuation on a \si{\deci\bel}-scale at \SI{90}{\degree}. The observed attenuations match this behavior closely, however, they are slightly smaller. Likely reasons for this are that the real vibration is not purely along one direction as already discussed in Sec.~\ref{sec:directionality}, and that signal components along other directions, e.g., noise, contribute to the total power as well.

\section{CONCLUSION} \label{sec:conclusion}

This work presented an analysis of the spectral and spatial characteristics of earbud vibrations induced by the wearer's \acl{ov}. Two earbud models were considered. Spectrally, the results confirm that \ac{ov}-induced earbud vibrations exhibit a low-pass characteristic, with high-power components below \SI{400}{\hertz} and an approximate \SI{-93}{\deci\bel} per decade roll-off towards higher frequencies. The exact cutoff varies with earbud model, likely a result of differences in coupling to the concha. One result of the pronounced low-pass characteristic is that multiple commercially available accelerometers are able to sense the high-power components at \ac{snr} above \SI{20}{\deci\bel} on average, but also that sensing frequencies above \SI{1}{\kilo\hertz} requires comparatively low-noise-floor accelerometers. This suggests that the high-power components, which are limited to \SI{400}{\hertz}, are the most relevant ones for many practical applications.

Spatially, the study shows that the high-power components of the vibration correspond mostly to an acceleration of the earbud in and out of the ear canal entrance. The vibration direction is largely consistent between subjects and fits, and similar for both earbud models considered. The consistency provides evidence that a single-axis accelerometer is sufficient for capturing these high-power components of the vibration. Simulations confirm this, with a favorably-mounted sensor resulting in a mean attenuation below \SI{1.5}{\deci\bel} of said components, and an unfavorably-mounted sensor resulting in \SI{11}{\deci\bel} or more. In contrast, capturing components around \SI{600}{\hertz} without attenuation requires a multi-axis sensor for one of the earbud models considered. Lastly, deviations of up to \SI{45}{\degree} from the optimal mounting direction result in less than \SI{3}{\deci\bel} additional attenuation of the high-power components. Overall, the results suggest that due to the spectral and spatial characteristics of the vibration, capturing the high-power components below \SI{400}{\hertz} with a single-axis sensor is likely a viable option for many applications.

\section{Acknowledgements}

The authors thank all participants of the measurements.

\newpage
\bibliographystyle{ieeetr}
\bibliography{literature.bib}

\end{document}